# Assembly Theory Explains and Quantifies the Emergence of Selection and Evolution

Abhishek Sharma,[1†] Dániel Czégel,[2,3†] Michael Lachmann,[5] Christopher P. Kempes,[4] Sara I. Walker,[2,4*] Leroy Cronin,[1*]

† Equal contribution [1] School of Chemistry, University of Glasgow, Glasgow, G12 8QQ, UK.

[2] BEYOND Center for Fundamental Concepts in Science, Arizona State University, Tempe, AZ, USA

[3] Institute of Evolution, Centre for Ecological Research, Budapest, Hungary

[4] School of Earth and Space Exploration, Arizona State University, Tempe, AZ, USA

[5] The Santa Fe Institute, Santa Fe, New Mexico, USA

*Corresponding authors' emails: Lee.Cronin@glasgow.ac.uk, sara.i.walker@asu.edu

**Since Darwin, scientists have struggled to reconcile the evolution of seemingly endless biological forms with a universe determined by fixed laws. The laws of physics underpin the origin of life, evolution, human culture and technology as set by the boundary conditions of the universe; however, they cannot predict the emergence of these things. By contrast evolutionary theory works in the opposite direction, indicating how selection can explain why some things exist and not others. To understand how open-ended forms can emerge in a forward-process from physics that does not include their design, a new approach to understanding and quantifying selection is necessary that applies across both physics and biology. Herein, we present a new theory, Assembly Theory (AT), which accomplishes this not by redefining the laws of physics, but instead by redefining the concept of an object that these laws operate on - in AT, objects are not considered as point particles, but instead are defined by the set of possible histories of their formation as an *intrinsic* property. Defined in this way, we show how objects themselves can provide *measurable* evidence of selection, whether within the well-defined boundary of an individual or selected unit, or not. We formalize by introducing a quantity called Assembly that captures how much causation was necessary to produce a given ensemble of objects. We demonstrate this theoretically and show how AT allows novelty generation and selection to enter naturally into the physics of complex objects by demonstrating how these can be described in a forward dynamical process accounting for the Assembly of objects. By redefining the concept of matter in terms of assembly spaces, AT provides a framework to unify descriptions of physics and biology in a new physics that emerges at the scale of chemistry where it is that history and causal contingency via selection starts to play a prominent role in what exists.**

**Introduction**

In evolutionary theory, selection[1] describes why some things exist and not others, but does so retroactively – by looking at what individuals survived.[2] This is not useful for explaining or predicting the emergence of novelty across chemistry, biology, and technology. Theories of physics do describe forward dynamics but achieve this by specifying an initial state[3] and evolving that state with a fixed set of laws. This means that novelty, i.e., things that are generated which are not predictable from the initial conditions, is not possible or must be introduced by randomness.[4] The open-ended generation of novelty[5] does not fit cleanly within either the paradigmatic frameworks of biology[6] or physics as laid out by Darwin[7] and Newton.[3]

There have been several efforts exploring the gap between physics and evolution,[8,9] such as dynamics where the laws are allowed to co-evolve with the states and considering the number of states to grow as a function of time. When the update rules are a function of the states, this results in systems that allow exploration of a larger space of possibilities,[10] but it does not solve the problem of where genuine novelty comes – the larger space still has a fixed size. Other efforts have removed the idea of a state entirely and studied the evolution of rules acting on other rules,[11] however, these models are so abstract it is difficult to see how they can describe – and predict – the evolution of physical objects. Yet another approach assumes a growing state space as a foundation and explores combinatorial spaces and their expansion, such as in the theory of the adjacent possible and related theories in statistical mechanics.[12] These approaches generates explosive growth in the number of possible configurations that cannot be sustained in a finite universe in finite time – these models generates novelty but do not include selection. Therefore, the approach lacks predictive power with respect to why only some evolutionary innovations happen and not others.

Herein, we introduce assembly theory (AT), which resolves these challenges by describing how novelty generation and selection can both operate in forward-evolving processes. The framework of AT allows us to not only predict features of novel forms in during selection, but also to quantify how much selection was necessary to produce the objects we have observed.[13,14] This can be done without having to pre-specify individuals or units of selection as selection is defined in terms of observed objects, their number of copies and how they can be built. In AT we do not need to get rid of the laws as formulated, or add ad hoc explosive expansion of states, instead, we re-invent the concept of an object that laws operate on. In AT, objects are not considered as point particles (as in most of physics) but instead are defined by the histories of their formation as an *intrinsic* property. We define this in terms of the assembly space. For a given object, the assembly space is defined as the pathways by which that object can be built from elementary building blocks, where we use only recursive operations combining what has already been built in the past. The assembly space captures minimal memory, in terms of the minimal number of physical operations necessary to assemble an observed object based on objects that could have existed in its past.[15]

We introduce the foundations of Assembly Theory (AT) and show how it can be implemented to quantify the degree of selection and evolution found in an ensemble of observed objects. We introduce a quantity, Assembly, built from two empirical observables: (1) the number of copies of the observed objects and (2) the objects' assembly indices, where assembly index is the minimal number of steps necessary to produce the object (its size). Assembly therefore quantifies the amount of memory necessary to produce a selected configuration of observed objects (with copy number and assembly index as coordinates) in a manner similar to how entropy quantifies the information necessary (or lack thereof) to specify the configuration (in terms of spatial positions and velocities as coordinates) of an ensemble of point particles. We demonstrate how AT leads to a unified language for describing selection and the generation of novelty. We thus show how AT provides a framework to unify descriptions of selection across physics and biology, with the potential to build a new physics that emerges at the scale of chemistry where history and causal contingency via selection start to play a prominent role in our interpretation of physical matter.

**Assembly Theory**

To quantify the emergence of selection and evolution we define what an object is in AT, and the key observable properties of objects so defined - (1) assembly index and (2) copy number. The concept of an object in Assembly Theory is simple and rigorously defined: an object is finite, distinguishable, persists over time, and the object is breakable such that the set of constraints to construct it from elementary building blocks are measurable. It should be noted that this definition of objects is in some sense opposite of standard physics, which treats objects as fundamental and unbreakable (e.g., the concept of 'atoms' as indivisible, which now applies to elementary particles). In assembly theory we recognize that the smallest unit of matter is typically defined by the limits of observational measurements and may not itself be fundamental, and that a more universal concept is to treat objects as anything that can be built. This allows us to naturally account for the emergent objects produced by evolution and selection as fundamental to the theory.

In principle, an object that exists in a large number of copies, allows the signatures describing the set of constraints that built it (the physical rules or laws governing its assembly) to be measured experimentally. For example, mass spectrometry can be used to measure the assembly for molecules, because it can measure the ways molecules are built by making bonds.[16] The concept of copy number is of fundamental importance in defining a theory that accounts for selection. This is because the more complex a given object is, the less likely it is that an identical copy of that object can exist without some precise mechanism that has been selected that generates that object. The consequences of this point will be explained in detail but simply put, complex objects that exist in high numbers indicate the presence of a physical system, itself the product of evolution and selection (e.g., a biological cell, technological factory, a human mind etc.), with a memory to build these objects.

**Assembly Index and Copy Number**

Assembly index is determined from the physical constraints that allow an object to be built. To construct an assembly pathway for an object, one starts from elementary building blocks composing that object and recursively joins these to form new structures, where at each recursive step the objects formed are added back to the assembly pool and are available for subsequent steps (see SI Section 1 and 2). For any given object $i$, we can define its assembly space as all recursively assembled pathways that produce it. For each object, the most important feature of its assembly space is the assembly index $a_i$, the size of the smallest assembly subspace that contains the object. This can be quantified as the length of the shortest assembly pathway which can generate the object, see Fig. 1.

In chemical systems, molecular assembly theory treats *bonds* as the elementary operations from which molecules are constructed. Thus, the shortest path to build the target molecule can be found by breaking the molecule into parts by breaking bonds, and then ordering those parts in order of size. The ordering is important since we start from the atoms, and add bonds, in sequence. Once we have generated a given motif on the path, this motif remains available for reuse. This recursive property is important because identifying the shortest recursive path allows us to quantify the minimum number of constraints, or memory size, to build the target molecule. Similarly, the assembly index can be estimated from any complex discrete object with well-defined building blocks, that can be broken apart, as shown in Fig. 1.

Assembly index can be read out directly from the target molecule, and as we will explain later, this is an intrinsic property that can be determined experimentally. This may seem like a small detail, but it is critically important for distinguishing assembly theory from complexity measures based on related but conceptually very different concepts of shortest paths and algorithmic compressibility that have emerged from computer science. For example, it is nowadays relatively straightforward to take the graph of a molecule and calculate approximations to its algorithmic complexity as the length of shortest program that can compute it.

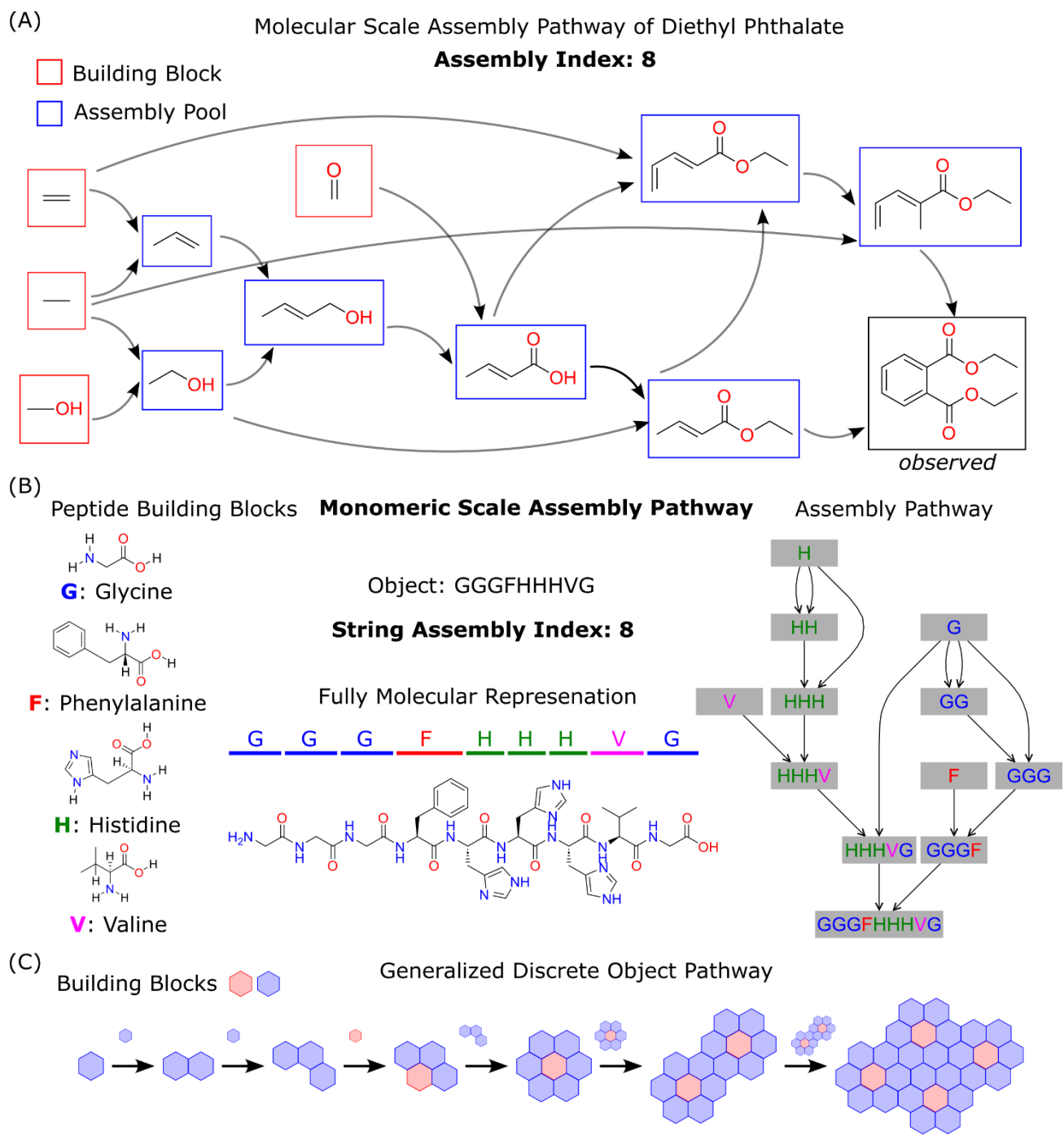

**Fig. 1 Assembly index and shortest path(s).** AT is general for all classes of objects – here illustrated with three different general types: (A) Assembly pathway to construct diethyl phthalate molecule considering molecular bonds as the building blocks. The figure shows the pathway starting with the irreducible constructs to create the molecule with assembly index 8. (B) Assembly pathway of a peptide chain by considering building blocks as strings. Left: shows four amino acids as building blocks, Middle: shows the actual object and its representation as a string, Right: Assembly Pathway to construct the string. (C) Generalized assembly pathway of an object which is comprised of discrete components.

However, these measures are features of the graphical representation of a molecule – and, more specifically the program used to compute the graph – and reflect nothing of the physics of how molecules are assembled in the real universe. Complexity measures derived from computer science[17–] [19] have no knowledge of chemistry, nor should we expect them to as they were not designed with that purpose in mind. As a simple example of where this could go wrong, one could build an "assembly space" with operations based on pasting atoms together (molecular alchemy!), instead of making bonds, and that would have a different graphical representation and minimal path, representing a different algorithm in computer science. And indeed, this could have a shorter minimal path than that found in the assembly space. However, this would not have a physical interpretation, because physical systems do not build molecules by combining atoms with no bonds – molecules are made by making bonds. Assembly theory is not an attempt to reinvent well-trodden concepts from computer science, but instead we aim to develop a new understanding of complex matter that naturally accounts for their selection and history in terms of what physical operations are permitted by the laws of physics.[20,21] We will discuss assembly theory as applied to molecules as the major application in this paper because their assembly index has been experimentally measured. In this case, assembly index has a clear physical interpretation and has been validated as quantifying evidence of selection in its application to the detection of molecular signatures of life. However, we expect the theory to be sufficiently general to apply to a wide variety of other systems including polymers, graphs, images, computer programs, and human languages as well as many others. The challenge in each case will be to construct an assembly space that has a clear physical meaning in terms of what operations can be caused to occur to make the object[20], see Fig. 1.

We also note that because assembly pathways are recursive, the process of constructing new objects retains the memory of the past formation of objects. This is important because the structure of assembly pathways implicitly implies two features of the environment the object is found in: (1) there are objects in its environment that can constrain the steps in the pathway and (2) these objects themselves have been selected because they must be retained over subsequent steps to physically instantiate the memory needed to build the target object. Among the most relatable examples of this are enzyme catalysts in biochemistry, which permit the formation of very unlikely molecules in large numbers, because the enzymes themselves are also selected to exist with a large number of copies. We make no distinction between the traditional notion of biological "individual" and objects that are selected in the environment to quantify the selection necessary to produce a given configuration. Thus, our approach naturally accounts for well-known phenomenon such as niche construction where organisms and environment are co-constructed and co-selected.

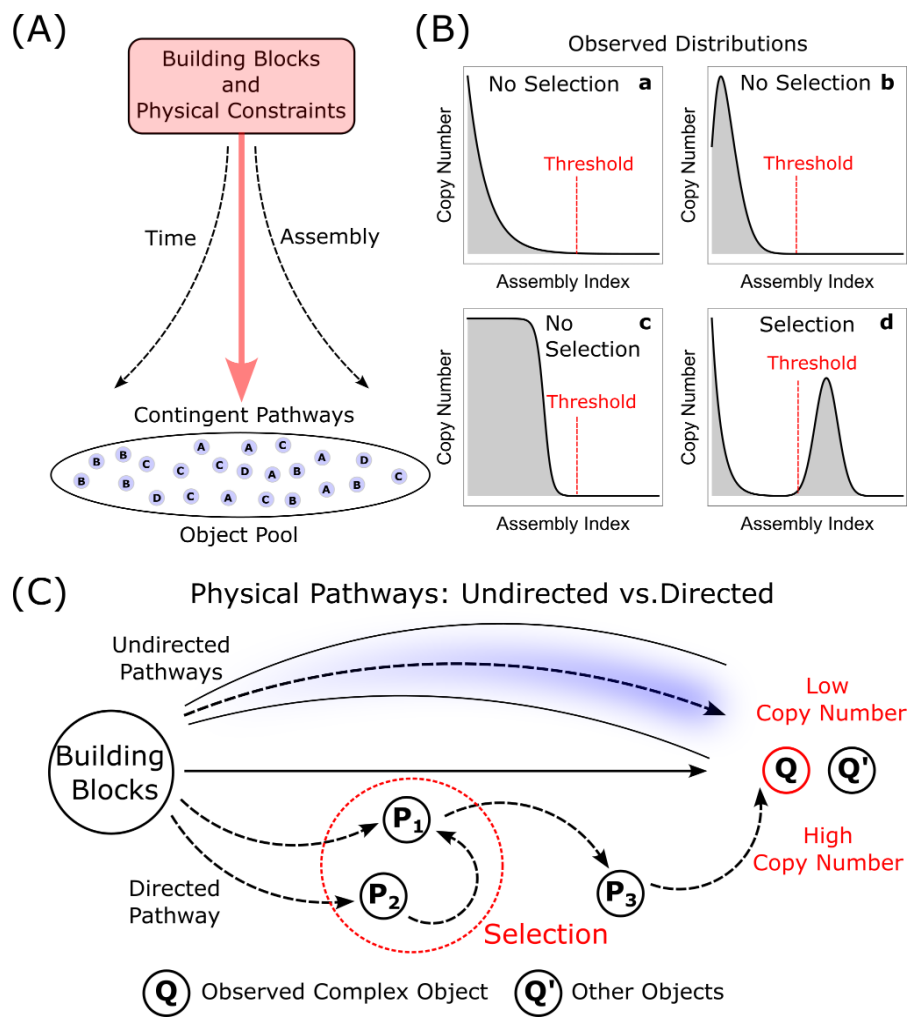

**Fig 2. Selection in Assembly Space.** (A) Pictorial representation of the assembly space represents the formation of combinatorial object space from building blocks and physical constraints. (B) Observed copy number distributions of objects at different assembly indices as an outcome of selection or no selection. (C) Representation of physical pathways to construct objects with undirected and directed pathways (selected) leading to the low and high copy numbers of the observed object.

Copy number is important since a single example of a highly complex molecule (with a very high assembly index) does not provide evidence of selection. A single object could potentially be generated in a series of random events that become increasingly less likely with increasing assembly index. If we consider a forward-building assembly process, without a specific target in mind, the number of possible objects that could be built at each recursive step grows super-exponentially in the absence of any constraints (see next section and SI Section 1). The likelihood to find and measure more than one copy of an object therefore decreases super-exponentially with assembly index in the absence of selection for a specific target. As such, finding more than one identical copy indicates the presence of a non-random process generating the object. Objects with high assembly index, found in abundance, provide evidence of selection because of the combinatorially growing space of possible objects at each recursive assembly step, see Fig 2. Notably, only objects with high copy numbers can have their assembly index measured since it is very hard technically to detect a given single instance of an 'object' and then infer its assembly index if only one of it exists. Thus, copy number and assembly index are intrinsically related. As an example, for the assembly index of molecules, the detection of identical molecules using a measurement technique like mass spectrometry requires the presence of many thousands of identical molecules.

**The Assembly Equation**

For a given ensemble of observed objects, we define Assembly as the empirical quantification of the total amount of selection that was necessary to produce that ensemble. We introduce the assembly equation (1):

$$A = \sum_{i=1}^{N} e^{a_i} \left( \frac{n_i - 1}{N_T} \right) \quad (1)$$

where $a_i$ is the assembly index of object $i$, $n_i$ is its copy number and $N_T$ as the total number of objects in the ensemble. We normalize by the number of objects in the ensemble such that Assembly between ensembles with different number of objects can be compared for quantifying processes such as selection. This is quantified within the summation by the distribution of the objects over the range of assembly indices. We note that empirical measurements of Assembly will constrain whether the associated constants in the Assembly Equation account for natural combinatorial explosion with no selection.

The assembly equation quantifies two competing effects (1) the complexity of discovering new objects and how (2) once discovered some objects become 'easier' to make because the memory has been selected for their formation. The exponential growth of Assembly with depth in the assembly space (quantified by assembly index) is derived by considering a linearly expanding assembly pool which has objects that combine at step $a_i \rightarrow a_i + 1$, where an object at the assembly index $a_i$ combines with another object from the assembly pool. Discovering new objects at increasing depth in an assembly space gets harder because the space of possibilities expands exponentially. Once discovered, the production of an object (copy number > 1) gets easier as the copy number increases because a high copy number implies that an object can be produced readily in a given context. Thus, the hardest innovation is making an object for the first time which is equivalent to the discovery, followed by making the first copy of that object, but once an object exists in very high abundance it must already be relatively easy to make, hence Assembly scales linearly with copy number for more than one object.

A consequence of the assembly equation is that increasing Assembly (*A*) results from increasing copy numbers *n* and increasing assembly indices *a*. If high assembly can be shown to capture cases where selection has occurred, it implies finding high assembly index objects in high abundance is a signature of selection. Along an assembly path, the information required at an assembly step to construct the object is "stored" within the object, see Fig. 2. When two objects are combined from an assembly pool the specificity of the combination process constitutes selection along an assembly path. As we will show, randomly combining objects within the assembly pool at every step does not constitute selection because no combinations exist in memory to be used again for building the same object again. If instead, certain combinations are preferentially used, it implies there must exist a mechanism that selects the specific operations and by extension specific target objects to be generated. We quantify the degree of selectivity as $\alpha$, which allows us to parameterize selection in an empirically observable manner (by parameterizing how much selection (reuses of specific sets of operations) went into generating a given ensemble). We demonstrate that the higher Assembly arises with the increase in selectivity. As an example, consider a simple forward process starting with fundamental particles, at each step a fraction of objects ($\tilde{\alpha}$) transforms into a higher assembly object $a_i \to a_{i+1}$ such that at assembly step t, the copy numbers $n_i(t)$ of objects with assembly index $a_i$ is given by $(1 - \tilde{\alpha}_i)n_i(t) + \tilde{\alpha}_{i-1}n_{i-1}(t)$ in the presence and the absence of the constraints. The estimated assembly (*A*) up to 50 assembly steps in shown in Fig. 3C and 3D.

We note that Assembly as given in Eq. (1) is determined for identified unique objects (with copy number > 1) and their assembly paths (we define the assembly path as the precise path of interactions that leads to the object). However, in real samples, there are almost always multiple coexisting, distinguishable objects, which in many cases include a common history for their formation. Transistors, for example, are used across multiple different technologies suggesting a common assembly subspace that includes transistor-like objects for many modern technologies. This common assembly space, constituting the overlap in the assembly path structure of two distinct structures is called a joint assembly space. For the generalization of the assembly equation that accounts for the joint assembly of objects, we introduce a version of the assembly equation that includes the quantification of shared pathways of the objects to determine the Assembly (*A*) of an ensemble (see SI Section 3).

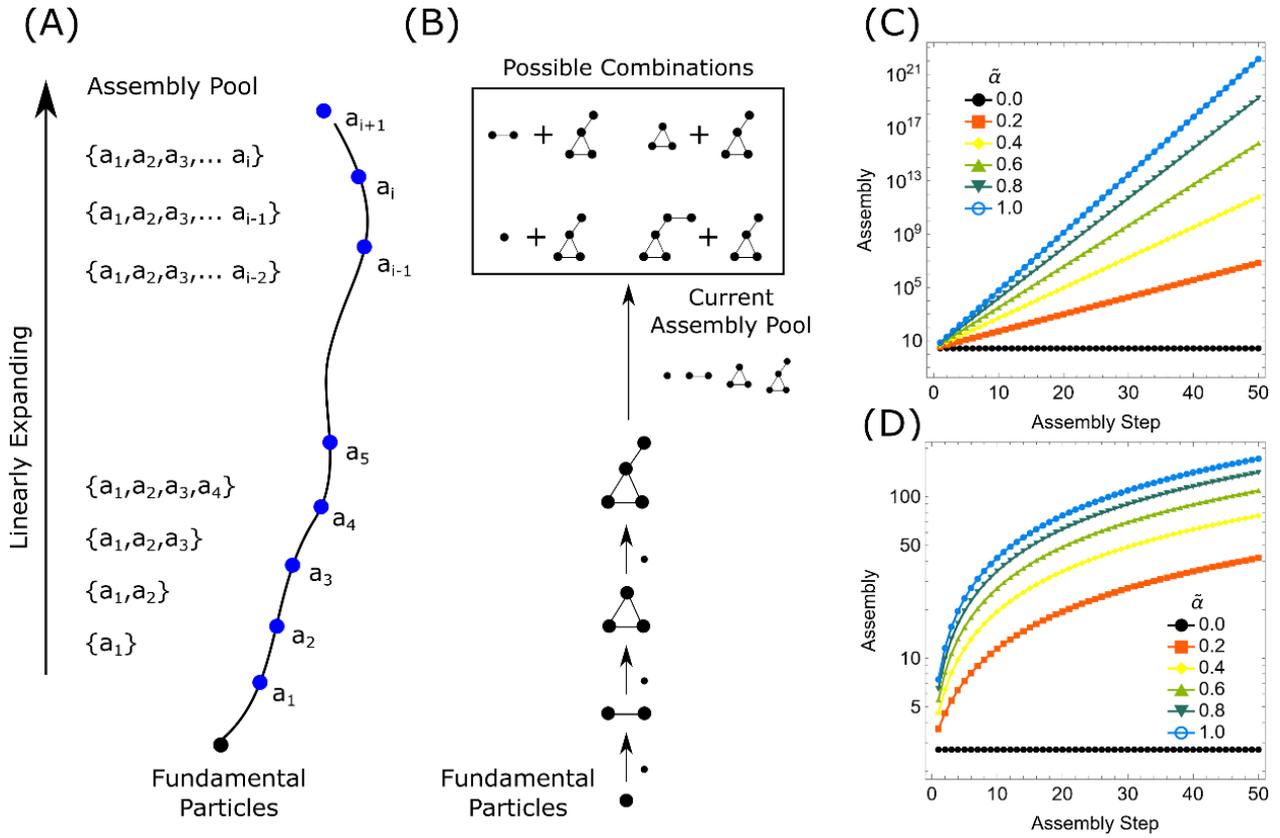

**Fig 3. Assembly of an ensemble.** (A) An isolated assembly path with a linearly expanding assembly pool, where at step $a_i \to a_i + 1$, $a_i$ objects are available for the next assembly step. (B) Linearly expanding assembly pool with adding one fundamental object at each step where the number of possible combinations at the $i^{th}$ step increases exponentially. (C) and (D) show the Assembly of the ensemble assuming isolated chains up to 50 assembly steps when $\tilde{\alpha}_i$ is constant with the assembly index and when $\tilde{\alpha}_i$ scales with assembly index $a$ as $\tilde{\alpha}_i = \tilde{\alpha}_0 f^a$, where $\tilde{\alpha}_0$ is fraction at the first step and $f \leq 1$ represents the increase in constraints with assembly steps. With the increase in $\alpha$ conversion fraction, objects at higher assembly emerge with higher copy numbers which for an isolated chain represents the efficiency of the forward process to construct complex objects.

## SELECTION WITHIN ASSEMBLY SPACES

The concept of the Assembly space allows the understanding of how selection and historical contingency imposes constraints on what can be made in the future. A key feature of assembly spaces is they are combinatorial because items are combined at every step. More objects exist in assembly space than can be built in finite time with finite resources because the space of possibilities grows super exponentially with the assembly index. Such combinatorial spaces do not play a prominent role in current physics, because current physics often considers objects as point particles and not as combinatorial objects (with limited exceptions). However, combinatorial objects are important in chemistry, and likewise for biology and technology, where most objects of interest (if not all) are hierarchical modular structures. While other work has explored the explosive (super-exponential) growth of combinatorial spaces to model the growth of novelty, these have so far not taken historical contingency into account. In assembly space, historical contingency is intrinsic, because the space is built compositionally, where items are combined recursively (accounting for hierarchical modularity). It is the combination of the compositionality with combinatorics that allows us to describe the selection, see Fig. 4.

To produce an assembly space, an observed object is broken down recursively to generate a set of elementary building units. These units can be used to then recursively construct the assembly pathways of the original object(s) to build what we call **Assembly Observed, $A_O$**. $A_O$ captures all histories for the construction of the observed object(s) from elementary building blocks, consistent with what physical operations are possible. Because objects in assembly theory are compositional, they contain information about the larger space of possible objects from which they were selected. To see how, we first build an assembly space from the same building blocks in $A_O$, which includes not just the pathways specific to assembling our original observed objects, but all possible pathways for assembling any object composed of the same elementary building blocks. The space so constructed is what we call the **Assembly Universe ($A_U$)**.

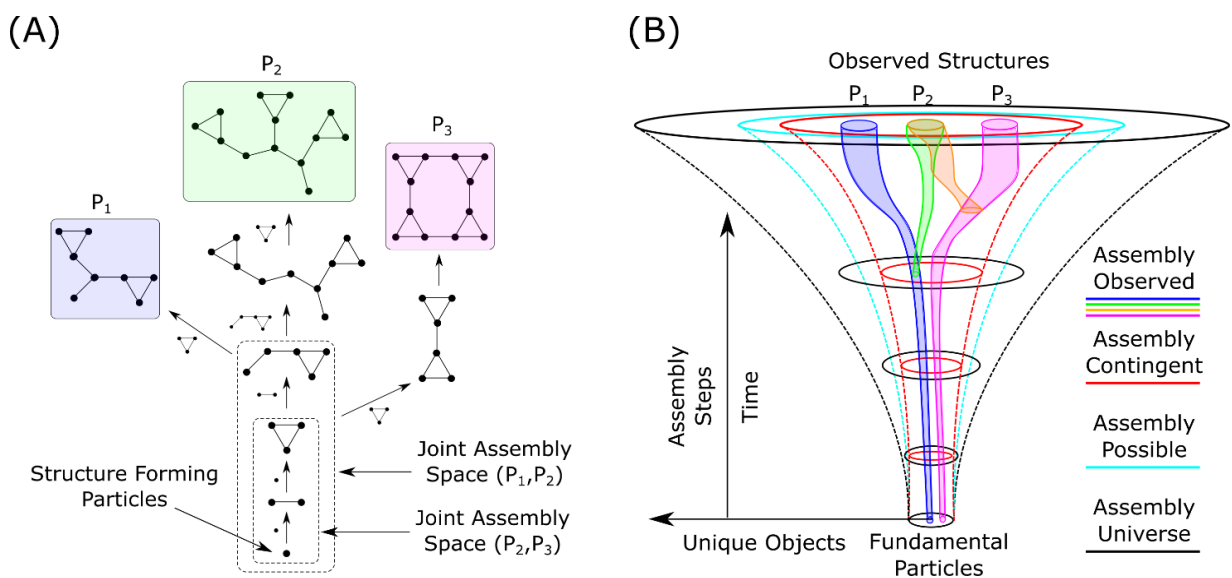

**Fig. 4 Assembly Spaces.** (A) Assembly Observed of the three objects shown as graphs ($P_1$, $P_2$, and $P_3$) with their shared minimal construction process called their Joint Assembly Space. (B) Illustration of the expansion of the Assembly Universe, Assembly Possible, Assembly Contingent, and Assembly Observed (see text for details). Note that the figure illustrates only their nested structure but not the relative size of the spaces where each set is typically exponentially larger than the subset.

In the **Assembly Universe ($A_U$)** all objects are possible with no rules, allowing a combinatorial explosion and hence the size of this space grows as a double exponential characteristic of exploding state spaces (adjacent possible), see SI Section 4 for details. While mathematically well-defined, this double-exponential growth is unphysical because the laws of physics place restrictions on what is possible (e.g., in the case of molecules an example is how quantum mechanics leads to specific numbers of bonds per atom). For most systems of interest, including in molecular assembly spaces, the assembly universe is orders of magnitude larger than the amount of matter available in the cosmologically observable universe, so there is no way to iteratively build and exhaust the entire space, even for relatively low assembly index objects, but for larger objects like proteins this can be truly gigantic[22]. Because everything can exist, there is an implication that objects can be constructed *independently* of what has existed in the past and of resource or time constraints, which is not what we observe in the real universe. We do not observe all possible objects at a given depth in the assembly space in the real universe because of selection. We next show how taking account of memory and resource limitation severely restricts the size of the space of what can be built, but also allows higher assembly objects to be built before exhausting resources constructing all the possible lower assembly objects. Assembly theory can account for selection precisely because of the historical contingency in the recursive construction of objects along assembly paths.

**Assembly Possible ($A_P$)** is the space of objects physically possible objects, which can be generated using the combinatorial expansion of the space using all the known rules of assembly for the objects (laws that govern how they can be built combinatorially) and allowing all the rules to be available at every step to every object. When an object with Assembly Index $a$ combines with its own history, its Assembly Index increases by one, $a \rightarrow a + 1$. If the resulting object can be made via other, shorter path(s), its Assembly Index will be smaller than $a + 1$. If the object combines with another which is not in its own history, the combination of them might have an Assembly Index that is larger than $a + 1$ (except if there is a shorter path as discussed above). These two effects may or may not statistically cancel. Another assumption behind the dynamical model of undirected dynamics is that it is microscopically driven by a stochastic rule that uses existing objects uniformly: the probability of choosing an object with Assembly Index $a$ to be combined with another is proportional to $N_a$, the number of objects in with Assembly Index $a$. We define the probability $P_a$ of an object being selected with Assembly Index ($a$) as $P_a \propto (N_a)^\alpha$, where $N_a$ is the number of objects with Assembly Index a. Here, $\alpha$ parameterizes the degree of selection: for $\alpha = 1$ all objects that have been assembled in the past are available for reuse, and for $0 \leq \alpha < 1$ only a subset (that grows non-linearly with assembly index) are available for reuse, indicating selection has occurred. This leads to the growth dynamics:

$$\frac{dN_{a+1}}{dt} = k_d(N_a)^\alpha \tag{3}$$

For $\alpha = 1$ there is historical dependence without selection. We build assembly paths by taking two randomly chosen objects from the assembly pool, combining them, and if a new object is formed adding this back to the pool. Here we are building random objects, but these are fundamentally different from random combinatorial objects because the randomness we implement is distributed across the recursive construction steps leading to an object. (See SI Section 5 for solutions). The case of $\alpha = 1$, where there is historical dependence, but no selection defines the boundary of Assembly Possible.

Within Assembly Possible, **Assembly Contingent ($A_C$)**, describes the shape of the space of objects formed where history, and importantly, selection on that history matters. Historical contingency is introduced by assuming only knowledge or constraints built on a given path or graph can be used in the future, or with different paths interacting in cases where selected objects that had not interacted previously now interact. Within Assembly Possible, the **Assembly Contingent ($A_C$)** can be found where $0 \leq \alpha < 1$ i.e. when selection can operate and the objects found in the space are controlled by a path-dependency contingent on each object that has already been built. The growth of the assembly selected is much slower growth than exponential; indeed, not all possible paths are explored equally, instead, the dynamics are channeled by constraints imposed by the selectivity emerging along specific paths. Indeed, a signature of selection in assembly spaces is a slower growth of unique objects than exponential growth. To show how, we use a simple linear polymer model as a phenomenological model to demonstrate how Assembly can different when selection happens. Starting with a single monomer in the assembly pool, the random exploration process combines two randomly selected polymers and adds them back to the assembly pool. In the case of selection, we implement here, the polymer that has been created most recently is selected to join a randomly selected polymer from the assembly pool. For both random construction and selection, this process was iterated up to $10^4$ steps and repeated 25 times. For each observed polymer in the assembly pool, the shortest pathway was generated (see SI Section 6 for details). For each run, the joint assembly space of all the observed polymers in the assembly pool was generated by the union of their shortest pathways, and the graph representation of the explored joint assembly space in undirected and directed exploration up to 100 steps is shown in Fig. 5 A and B. To quantify the degree of exploration at a given assembly step, we calculate the exploration ratio which is defined by the ratio of observed nodes and the total number of nodes present in the joint assembly space. Fig. 5C shows the exploration ratio and the mean maximum assembly index observed (approximated by $\log_2(n)$ where $n$ is the length of the polymer) for the undirected and directed exploration processes. Comparing the directed process to the undirected exploration illustrates a central principle: the signal of selection is simply a lower exploration ratio and higher complexity (as defined by the mean assembly index). The observation of lower exploration ratio in the directed process as compared to the undirected process is the evidence of the presence of the selectivity in the combination process between the polymers existing in the assembly pool. Here, the directed process aims to produce longer chains by sorting and selecting longer chains from the assembly pool for combination restricting the exploration. This is also evident observation of polymers with higher assembly index in the case of directed process. The process representing sorting and selecting chains within the assembly pool represents an outcome of a physical process leading to selection (see SI Section 7 for an additional model).

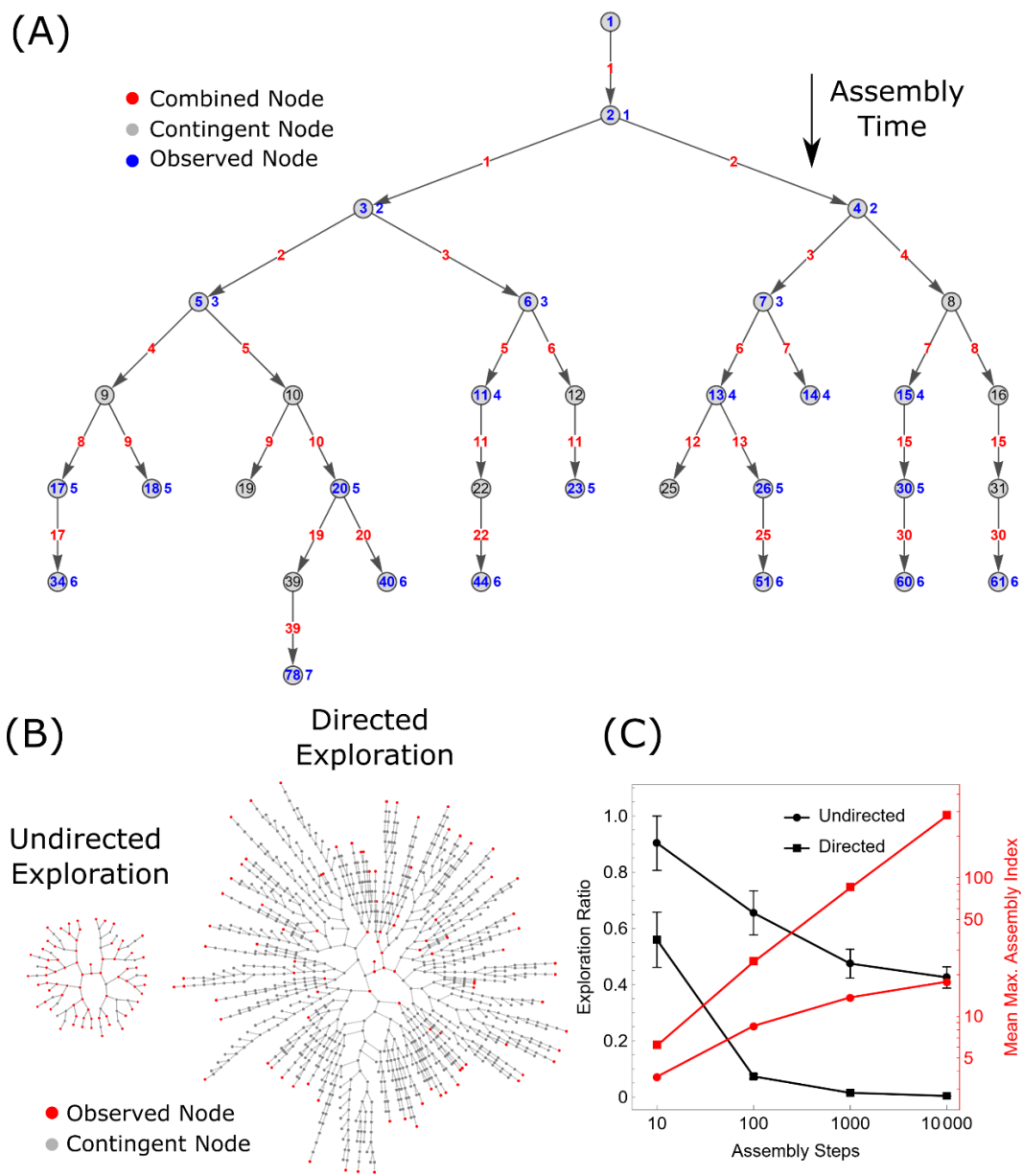

**Fig. 5 Undirected and directed exploration in a forward assembly process.** (A) The joint assembly space of polymers (with their lengths indicated) after 30 steps created by combining randomly selected polymers from the assembly pool. Realized polymers are shown in blue ("observed nodes"), whereas grey nodes represent polymers that have not been realized but are part of the joint assembly space of all realized objects. Polymers with lengths indicated within the nodes are randomly combined with ones indicated as red edge labels to yield a new polymer. (B) The comparison between undirected and directed exploration after 100 assembly steps using a graph with radial embedding. (C) The exploration ratio (defined by the ratio of the number of observed nodes and total nodes which includes observed and causal nodes) and mean maximum assembly index (averaged over 25 runs) up to $10^4$ assembly steps.

Our conjecture is that the 'more assembled' an object, the more selection is required for it to come into existence. The historical contingency in assembly spaces means that assembly dynamics explores higher-assembly objects before exhausting all lower-assembly objects, leading to a vast separation in scales between the number of objects that could be explored versus those that are actually constructed following a particular path. This early symmetry breaking along historically contingent paths is a fundamental property of all assembly processes. It also introduces an "assembly time" that ticks at each object being made: assembly physics includes an explicit arrow of time intrinsic to the structure of assembly spaces.

**ASSEMBLY UNIFIES THE EMERGENCE OF SELECTION WITH PHYSICS**

In the real universe, objects can only be built from parts that already exist. The discovery of new objects is therefore historically contingent. The rate of discovery of new objects can be defined by the expansion rate ($k_d$) introduced in the last section, which describes the rate of formation of unique objects in the assembly space from previously existing objects. This introduces a characteristic timescale $\tau_d$, we define as the discovery time scale. Additionally, once an object is discovered it can be reproduced if there exists a mechanism in the environment selected to build it again. Selected objects, therefore, increase in copy number, and the copy numbers for these objects must obey mass transfer kinetics. Thus far, we have considered discovery dynamics within the assembly spaces which do not account for the abundance or copy number of the observed objects once they are discovered. To include copy number in the dynamics of assembly spaces we must introduce a second timescale, the rate of production ($k_p$) of a specific object with a characteristic time scale $\tau_p$ (production time scale), see Fig. 6. For simplicity, we assume the selectivity and interaction among the emerging objects in assembly space is similar across objects. Defining the two separate timescales for (1) initial discovery of an object ($k_d$), and (2) making copies of existing objects ($k_p$) allows us to determine the regimes where selection is possible, see Fig 6.

For $\frac{k_d}{k_p} \gg 1$, where objects are discovered quickly but reproduced slowly, the expansion of assembly space is too fast under mass constraints to accumulate a high abundance of any given objects, and this leads to a combinatorial explosion of unique objects with low copy numbers. This, for example, is consistent with how some unconstrained prebiotic synthesis reactions, such as the formose reaction end up producing tar, which is composed of a large number of unidentifiable molecules because they have low copy number.[23] Selection and evolution cannot emerge if new objects are generated on timescales so fast that resources are not available for making more copies of objects that already exist.

For $\frac{k_p}{k_d} \ll 1$, objects are reproduced quickly but new ones are discovered slowly. Here resources are primarily consumed in producing additional copies of objects that already exist. New objects are discovered infrequently. This leads to a high abundance of objects with low assembly. Significant separation of the two timescales of discovery of new objects and (re)production of selected objects, results in either a combinatorial explosion of objects with low copy numbers, or conversely, high copy numbers of low assembly objects. In both cases, we will not observe trajectories that grow more complex structures.

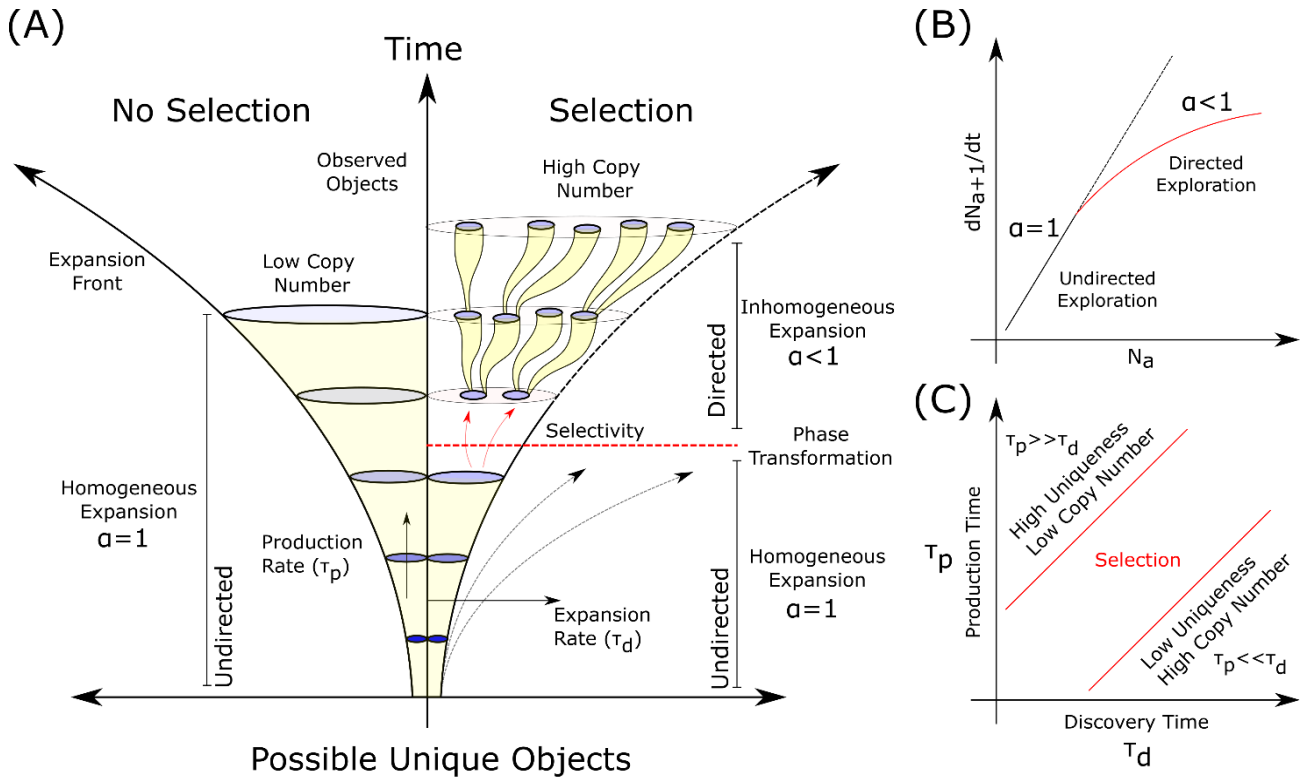

**Fig. 6 Selection and evolution in Assembly Space.** Assembly processes with and without selection. The selection process is defined by a transition from undirected to directed exploration. The parameter $\alpha$ represents the selectivity of the assembly process ($\alpha = 1$: undirected/random expansion, $\alpha < 1$: directed expansion). Undirected exploration leads to the fast homogeneous expansion of discovered objects in the assembly space, while directed exploration leads to a process that is more like a depth-first search. Here, $\tau_d$ is the characteristic timescale of discovery, determining the growth of the expansion front, while $\tau_p$ is the characteristic timescale of production that determines the rate of formation of objects (increasing copy number). **(B)** Rate of discovery of unique objects at assembly $a + 1$ vs. number of objects at assembly $a$. The transition of $\alpha = 1$ to $\alpha < 1$ represents the emergence of selectivity limiting the discovery of new objects. **(C)** Phase space defined by the discovery ($\tau_p$) and production ($\tau_d$) time scale. The figure shows three different regimes: (1) $\tau_d \ll \tau_p$, (2) $\tau_d \gg \tau_p$, and (3) $\tau_d \approx \tau_p$. Selection is unlikely to emerge in regimes 1 and 2, and possible in regime 3.

Therefore, the emergence of selection and open-ended evolution in a physical system would be indicated by a transition from $k_p < k_d$ (undirected exploration to create unique objects) to $k_d < k_p$ (selective dynamics to sustain copy number), see Fig. 6, where there cannot be a large separation in the timescales between discovering new objects and reproducing ones that are selected.

Our goal in studying mass action kinetics using assembly theory to describe objects is twofold: (1) we aim to show how the generation of novelty can be described alongside selection in a forward-process (thus unifying key features of life with physics) and (2) we aim to show that in the regime where selection is operative, measuring the Assembly (defined in Eq. 1) identifies how much selection occurred. We can accomplish both by studying phenomenological models, with the understanding we are putting selection in by hand here to demonstrate the principles of how the Assembly equation quantifies selection so that it can be applied with confidence to cases where we may not know whether or not selection was operative.

Consider a forward assembly process where the copy number of the emerging species follows homogeneous kinetics, together with the discovery dynamics as given by equation 3. With the discovery of new unique objects over time, symmetry breaking in the construction of assembly paths will create a network of growing branches within the assembly space. Consider a simple forward assembly process where the concentration of copy number of a single branch is represented by $n_1(t) \to n_{2,i}(t) \to n_{3,j}(t) \to \cdots n_{a,p}(t) \to n_{a+1,q}(t) \ldots$ where, $n_{a,p}(t)$ is the copy number of the unique objects at assembly $a$ with index $p$ (likewise defined for assembly $a+1$ with index $q$). We assume the production rate constants for producing objects (independent of $p$ and $q$) decrease with increasing assembly index by a factor $\beta$ at each step, with $k_f$ the initial rate constant at the first step, hence $k_p(a) = k_f \beta^a$. This is a phenomenological choice, motivated by the observation that with increasing assembly index, the pathways for the formation of the object have more steps and must be more specific: both of these effects lead to a decrease in the production rate relative to less assembled objects. As a further simplification, we assume all objects at the assembly index $a$ have the same copy number, such that the total number of objects at the assembly $a$ at a given time $t$ is given by $N_a(t)n_a(t)$. It is important to note that this model is a simple continuum description of the physical phenomena of discovery and mass transfer kinetics and because it is continuous it can lead to unphysical copy number values of less than one (see SI Section 8 for the complete mathematical description of the model).

Fig. 7A-E shows the distribution of copy numbers of unique objects ($n_a(t)$) up to the dimensionless time $10^9$ for assembly indices <=25 and for selectivity parameter $\alpha = 0.01, 0.2, 0.5, 0.8, 1.0$. With the increase in $\alpha$ (corresponding to a *decrease* in selectivity) the number of unique objects increases rapidly, leading to a rapid decline in the respective copy numbers of objects over time (Fig. 7A-E). The time of unique discovery of an object with copy number 1 increases with the assembly index, which is governed by the forward process, see Fig. 7.

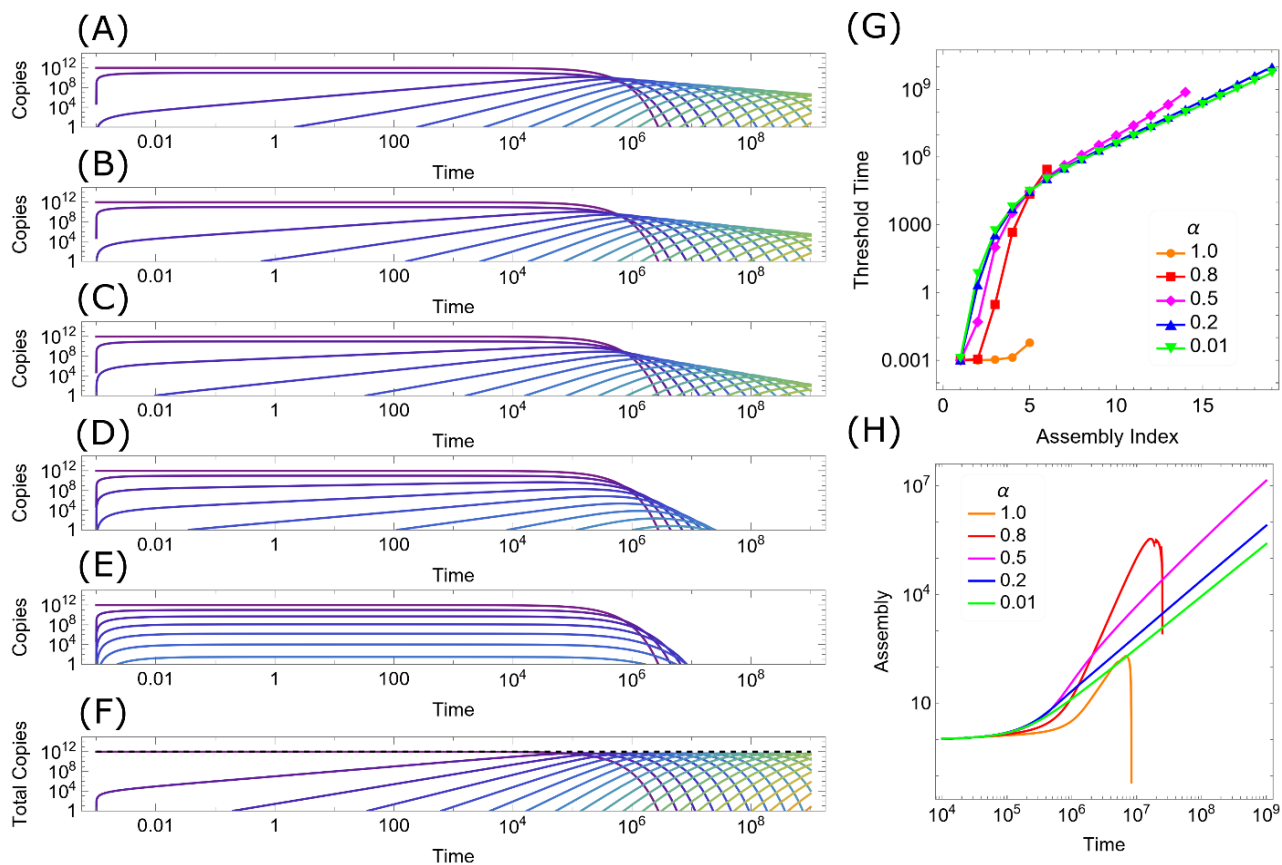

**Fig 7. Dynamics in assembly contingent coupled with kinetics.** (A)-(E) Observed Copy numbers of unique objects for assembly index 0-25 with $\alpha = 0.01, 0.2, 0.5, 0.8, 1.0$. (F) The total number of objects at different assembly indices, which will be equal for all cases independent of $\alpha$. (See SI Section 8 for details). (G) Threshold time required to reach copy number: 10 copies of objects at different assembly indices at various $\alpha$ values. (H) Assembly of the ensembles vs. time to generate the ensembles at various $\alpha$ values.

A key feature of assembly theory is that objects must be observable – that is they should exist in more than one copy. Fig. 7G shows the discovery time defined as the **Threshold time** when the copy number of an object reaches a minimum threshold for detection (in this case we set that value to 10 copies to illustrate how this works). For a mass spectrometer, this minimal threshold is ca. 10,000 copies. The threshold time depends on the limitations of measurement (the threshold for detection of the object), the discovery and production time scales and the selectivity $\alpha$. As an illustrative example, consider a system with fixed selectivity: a fast discovery rate means new objects can be discovered more rapidly, but a slower production rate would have the effect that it takes longer to achieve the minimal threshold copy number for detection. By contrast, with a slower discovery rate, it would take longer to discover an object, however that could be compensated by a faster production rate where it takes much less time to achieve the threshold copy number once an object is discovered for the first time.

To show how Assembly captures when selection has driven the generation of ensembles of high complexity, we calculated A for ensembles with varying selectivity $\alpha$ as shown in Fig. 7H. For a physical process with no selection ($\alpha \approx 1$), the production process is not sustainable and copy numbers decrease rapidly leading to a fall in Assembly over time. With an increase in selectivity ($\alpha < 1$), the number of unique products is restricted, and the production process is sustainable over higher assembly values as high copy numbers are produced. On the opposite extreme, for the very selected and over-constrained processes ($\alpha \approx 0$) few unique products are produced with a very high copy number, and the Assembly is low because of the low diversity of objects and low assembly indices.

The interplay between the two characteristic time scales describes discovery dynamics ($\tau_d \sim 1/k_d$) and forward kinetics ($\tau_p \sim 1/k_p$) together with selection (characterized by the selectivity parameter $\alpha$) are essential for driving processes towards creating higher assembly objects. This is characteristic of trajectories within Assembly Contingent. Assembly captures key features of how the open-ended growth of complexity can only occur within this restricted space by generating new objects with increasing assembly, while also producing them with a high copy number. Restricted selectivity ($\alpha < 1$) together with comparable time scales ($\tau_d \approx \tau_p$) are essential characteristics necessary to produce high Assembly ensembles. This is quantified by the Assembly equation which quantifies the degree of selectivity that must exist for an observed ensemble, and both the assembly index and copy number can be both computed and experimentally measured. The selectivity within an unknown physical process can be elucidated by experimentally detecting the number of species, their assembly index, and copy number as a function of time. Assuming that the species observed using analytical techniques such as mass spectrometry have a high copy number, the discovery rate and the selectivity coefficient ($\alpha$) can be computed using the rate of formation of new species and current existing species at all observed assembly indices.

**Conclusions**

We have introduced the foundations of assembly theory and how it can be implemented to quantify the degree of selection and evolution found in an ensemble of objects, agnostic to the detailed formation mechanisms of the objects or knowing a priori which objects are products of selection (e.g., are units of selections or individuals in biology). To do so, we introduce a quantity, Assembly, built from two empirical observables of assembly spaces: (1) the number of copies of an object and (2) its assembly index, where the assembly index is the minimal number of recursive steps necessary to build the object (its size). We demonstrate how AT allows a unified language for describing selection and the generation of novelty by showing how it quantifies the generation of novelty and selection at the same time in a forward process describing the mass-action kinetics of an ensemble of assembled objects. We thus show how AT provides a framework to unify descriptions of selection across physics and biology, with the potential to build a new physics that emerges in chemistry where history and causal contingency via selection must start to play a prominent role in our descriptions of matter. For molecules, computing the assembly index is not explicitly necessary, because the assembly index can be probed directly experimentally with high accuracy with spectroscopy techniques including mass spectroscopy, IR, and NMR spectroscopy[24].

## Author Contributions

LC and SIW conceived the theoretical framework building on the concept of the theory. AS and DC developed the mathematical basis for the framework, explored the assembly equation, and AS did the simulations with input from DC. ML and CK helped with the development of the fundamentals of assembly theory. LC and SIW wrote the manuscript with input from all the authors.

## Acknowledgements

LC, SIW, DC and AS would like to acknowledge our teams and colleagues at the University of Glasgow and Arizona State University for discussions including Cole Mathis, Douglas Moore, Stuart Marshall, and Paul Davies. We acknowledge financial support from the John Templeton Foundation (grants 61184 and 62231), EPSRC (grant nos. EP/L023652/1, EP/R01308X/1, EP/S019472/1, and EP/P00153X/1), the Break-through Prize Foundation and NASA (Agnostic Biosignatures award no. 80NSSC18K1140), MINECO (project CTQ2017-87392-P), and ERC (project 670467 SMART-POM).

## References

1. Kauffman, S. A. *The origins of order: self-organization and selection in evolution*. (Oxford University Press, 1993).

2. Gregory, T. R. Understanding Natural Selection: Essential Concepts and Common Misconceptions. *Evol. Educ. Outreach* **2**, 156–175 (2009).

3. Newton, I. *Newton's Principia. The Mathematical Principles of Natural Philosophy. (Daniel Adee)*. (1846).

4. Cross, M. C. & Hohenberg, P. C. Pattern formation outside of equilibrium. *Rev. Mod. Phys.* **65**, 851–1112 (1993).

5. Carroll, S. B. Chance and necessity: the evolution of morphological complexity and diversity. *Nature* **409**, 1102–1109 (2001).

6. Chesson, P. Mechanisms of Maintenance of Species Diversity. *Annu. Rev. Ecol. Syst.* **31**, 343–366 (2000).

7. Darwin, C. *On the origin of species by means of natural selection, or, The preservation of favoured races in the struggle for life*. (Natural History Museum, 2019).

8. Tilman, D. *Resource Competition and Community Structure. (MPB-17), Volume 17*. (Princeton University Press, 2020). doi:10.2307/j.ctvx5wb72.

9. Elena, S. F., Cooper, V. S. & Lenski, R. E. Punctuated Evolution Caused by Selection of Rare Beneficial Mutations. *Science* **272**, 1802–1804 (1996).

10. Lutz, E. Power-Law Tail Distributions and Nonergodicity. *Phys. Rev. Lett.* **93**, 190602 (2004).

11. Fontana, W. & Buss, L. W. The barrier of objects: From dynamical systems to bounded organizations. in *Boundaries and Barriers. J. Casti and A. Karlqvist (eds.)* 56–116 (Addison-Wesley, 1996).

12. Cortês, M., Kauffman, S. A., Liddle, A. R. & Smolin, L. The TAP equation: evaluating combinatorial innovation in Biocosmology. Preprint at http://arxiv.org/abs/2204.14115 (2023).

13. Marshall, S. M., Murray, A. R. G. & Cronin, L. A probabilistic framework for identifying biosignatures using Pathway Complexity. *Philos. Trans. R. Soc. Math. Phys. Eng. Sci.* **375**, 20160342 (2017).

14. Marshall, S. M., Moore, D. G., Murray, A. R. G., Walker, S. I. & Cronin, L. Formalising the Pathways to Life Using Assembly Spaces. *Entropy* **24**, 884 (2022).

15. Liu, Y. *et al.* Exploring and mapping chemical space with molecular assembly trees. *Sci. Adv.* **7**, eabj2465 (2021).

16. Marshall, S. M. *et al.* Identifying molecules as biosignatures with assembly theory and mass spectrometry. *Nat. Commun.* **12**, 3033 (2021).

17. Arora, S. & Barak, B. *Computational Complexity: A Modern Approach*. (Cambridge University Press, 2009). doi:10.1017/CBO9780511804090.

18. Wallace, C. S. Minimum Message Length and Kolmogorov Complexity. *Comput. J.* **42**, 270–283 (1999).

19. Bennett, C. H. Logical Depth and Physical Complexity. in *The universal Turing machine, a half century survey* 227–257 (Oxford University Press, 1988).

20. Deutsch, D. & Marletto, C. Constructor theory of information. *Proc. R. Soc. Math. Phys. Eng. Sci.* **471**, 20140540 (2015).

21. Marletto, C. Constructor theory of life. *J. R. Soc. Interface* **12**, 20141226 (2015).

22. Beasley, J. R. & Hecht, M. H. Protein Design: The Choice of de Novo Sequences. *J. Biol. Chem.* **272**, 2031–2034 (1997).

23. Kim, H.-J. *et al.* Synthesis of Carbohydrates in Mineral-Guided Prebiotic Cycles. *J. Am. Chem. Soc.* **133**, 9457–9468 (2011).

24. Jirasek, M. *et al.* Multimodal Techniques for Detecting Alien Life using Assembly Theory and Spectroscopy. (2023) doi:10.48550/ARXIV.2302.13753.